\documentclass{moriond}

\usepackage{amsmath,amssymb}

\def\be{\begin{equation}}
\def\ee{\end{equation}}
\def\bea{\begin{eqnarray}}
\def\eea{\end{eqnarray}}

\newcommand{\Photo}{}

\begin{document}
\vspace*{4cm}
\title{FLAVOR HIERARCHIES AND B-ANOMALIES FROM 5D}

\author{J. M. LIZANA}

\address{Physik-Institut, Universit\"at Z\"urich, Winterthurerstrasse 190, \\
CH-8057 Z\"urich, Switzerland}

\maketitle\abstracts{
$B$-anomalies may suggest New Physics at the TeV scale breaking flavor universality. In particular, 4321 gauge models can successfully explain them in a consistent way. In this talk we explore how to UV complete the 4321 model in a 5D warped background to solve simultaneously the Higgs hierarchy problem too, finding interesting connections with the flavor puzzle. We present a model that addresses the $B$-anomalies, flavor hierarchies, and the Higgs hierarchy problem, where quarks and leptons are unified à la Pati-Salam in a non-universal way, and the Higgs appears as a pseudo-Nambu-Goldstone boson. These proceedings are based on arXiv:2203.01952.
}

\section{Introduction}

Flavor is an intriguing feature of the Standard Model (SM). It is an exact symmetry of the gauge sector only broken by the Yukawa couplings, which have a very particular hierarchy. Although the explanation of this structure could be postponed to some very high scale of New Physics, the observation of the $B$-anomalies\,\cite{Cornella:2021sby} in the recent years may suggest hints of New Physics connected to a flavor hierarchy explanation.\cite{Bordone:2017bld}

$B$-anomalies are deviations in the $B$-meson semileptonic decays that break lepton flavor universality, both, in neutral current $b\to sll$, and charge current $b \to c\tau \nu$ processes. The neutral current anomalies point to a New Physics effective scale $\sim 40\,{\rm TeV}$, while for the charge current ones, the effective scale is much lower $\sim 3\,{\rm TeV}$.
The difference on these scales, together with the difference on the family number of the fields involved in each process ($3_q\to 2_l 2_l 2_q$ versus $3_q\to 3_l 3_l 2_q$) suggests that a combined explanation should be mainly coupled to the third family. If we assume that the New Physics has a suppression $\epsilon_q,\epsilon_l \sim 0.1$ when it is coupled to the second family, both set of anomalies point to the same scale around the TeV.
This structure allows to design models with an approximate $U(2)$ flavor symmetry in the light families, minimally broken to reproduce the SM Yukawas. This is useful to protect a model from stringent bounds from flavor-physics observables in the light families, while keeping a TeV New-Physics scale as needed for the $B$-anomalies.

\section{4321 model and beyond}

The most promising single mediator that can account for both sets of $B$-anomalies is the $U_1$ vector leptoquark,\cite{Angelescu:2021lln} with quantum numbers $({\bf 3},{\bf 1})_{2/3}$ under the SM gauge group $SU(3)_c\times SU(2)_L\times U(1)_Y$. The interaction terms with the SM fields we can write up to dimension 4 are
\begin{equation}
\mathcal{L} \ni \frac{g_U}{\sqrt{2}}U_1^{\mu}\left[ \beta_L^{i\alpha} (\bar q_L^{\,i} \gamma_{\mu}l_L^{\alpha}+\beta_R^{i\alpha} (\bar d_R^{\,i} \gamma_{\mu}e_R^{\alpha})\right] +h.c.\label{eq:U1couplSM}
\end{equation}
where we can choose the basis where the down and right-handed (RH) fields are mass eigenstates. A good $B$-anomaly explanation consistent with other flavor observables requires 
$|\beta_L^{i \alpha}| \sim \epsilon_q^{n_{i}} \epsilon_l^{n_{\alpha}}$, where $n_{i,\alpha}$ is $0$ for the third family, and increases in one unit each time the family number is decreased. The RH couplings can only be non-highly suppressed for $\beta_R^{b\tau}$, to forbid dangerous contributions to some flavor observables as $B_s\to \mu\mu$.\cite{Cornella:2021sby}

Such vector field appears as massive gauge boson in the so called 4321 models,\cite{Bordone:2017bld,Cornella:2019hct,Greljo:2018tuh} based on the gauge group $SU(4)_h\times SU(3)_l \times SU(2)_L\times U(1)_X$.
Here, $h$ and $l$ denotes the third and light families respectively. Only the third family is charged under $SU(4)_h$, so third-family quarks and leptons are unified à la Pati-Salam, and only the first and second families are charged under $SU(3)_l$. The abelian component of the group is $X=B_l-L_l+T^3_R$, where $B_l$, $L_l$ are baryon and lepton number of the light families, and $T^3_R$ is the RH isospin. This group breaks flavor universality in the gauge sector, keeping a $U(2)$ symmetry in the light families. If the 4321 gauge symmetry is broken at the TeV scale to the SM gauge group, a massive $U_1$ coupled to the third family appears, as required to explain the $B$-anomalies in first approximation, together with a massive $Z^{\prime}$ and a color octet $G^{\prime}$. These extra massive vector bosons do not affect the $U_1$ explanation of the $B$-anomalies but they provide interesting connections with other observables in colliders and flavor physics.\cite{Cornella:2021sby} 

To generate the light-heavy Yukawa mixing, and couplings of the leptoquark with the second family as required in Eq.~\eqref{eq:U1couplSM}, we need to include a multiplet of vector-like fermions charged under $SU(4)$ with the same quantum numbers as the third family $SU(2)_L$-doublets. They can mix with an $O(\epsilon_{q,l})$ angle with the light families after 4321 breaking, inducing a minimal breaking of the $U(2)$ flavor symmetry in the left-handed sector that generates the required couplings both in the Yukawa and leptoquark sectors. Additional breakings of $U(2)$ in the RH sector would be dangerous for flavor observables as $\Delta F=2$ processes or $B_s\to \mu \mu$, so RH light family fields cannot mix with $SU(4)$-charged fields. This is naturally forbidden by gauge symmetry in the 4321 model with the described matter content.

\subsection{Warped extra dimensions and multi-scale origin of the flavor hierarchies}

New Physics at the TeV scale invites to look for combined explanations with the Higgs hierarchy problem. One of the most popular proposals to stabilize the scale of the Higgs potential are Randall-Sundrum (RS) models,\cite{Randall:1999ee} or their dual 4D picture: strongly coupled dynamics where the Higgs is a composite state. RS models are built on warped 5D spacetime with a compact extra dimension between two 4D branes, with a geometry
\begin{equation}
{\rm d}s^{2} = e^{-2k y}\, \eta_{\mu\nu}\, {\rm d}x^{\mu} {\rm d}x^{\nu} - {\rm d}y^{2}.\label{RSgeometry}
\end{equation}
Here, $k$ is the curvature scale, and the branes, called ultraviolet (UV) and infrared (IR) branes, are localized at different values of the extra dimension $y$, $y_{\rm UV}<y_{\rm IR}$.
The warping factor produces a gravitational redshift effect that links the position in the extra dimension to the energy scale, $\Lambda \sim e^{k(y-y_{\rm IR})}\Lambda_{\rm IR} $, where $\Lambda_{\rm IR}=ke^{-ky_{\rm IR}}$ is the confining scale of the dual strongly coupled theory, or the Kaluza Klein (KK) scale of the 5D theory. Lower values of $y$ are associated to higher scales, so the Higgs must be localized in the IR brane to solve the large Higgs hierarchy problem.

The most popular approach to flavor in these models is the anarchic partial compositeness paradigm.\cite{Gherghetta:2000qt} The hierarchies of masses and mixing angles are achieved due to the different exponential suppression of the fermion zero modes of the SM fields in the IR brane, where the Higgs is located.
However, although these models have an effective $U(2)$ flavor protection at the KK scale from a GIM-like mechanism in RS,\cite{Gherghetta:2000qt} the breaking of the $U(2)$ symmetry is not minimal in general. There may be additional sources of $U(2)$-breaking in the RH sector, which are specially dangerous if we want to embed a 4321 model in a RS geometry. 

A possible way to improve this situation is to generate the Yukawa couplings of the three families at three different scales, i.e. different positions in the extra dimension $y$.\cite{Panico:2016ull,Fuentes-Martin:2020pww,Fuentes-Martin:2022xnb} This can be realized introducing extra intermediate branes where the different families are localized. The Higgs vacuum expectation value (VEV) profile has a decreasing profile towards the UV, so higher scales will suppress the resulting Yukawa. In this picture, RH fields can be highly, or completely localized on the branes, so the only breaking of the $U(2)$ symmetry of the light RH fields will be the suppressed light Yukawa couplings, protecting flavor bounds.

\section{A 5D model}

\begin{figure}
\centering
\includegraphics[width=0.75\textwidth]{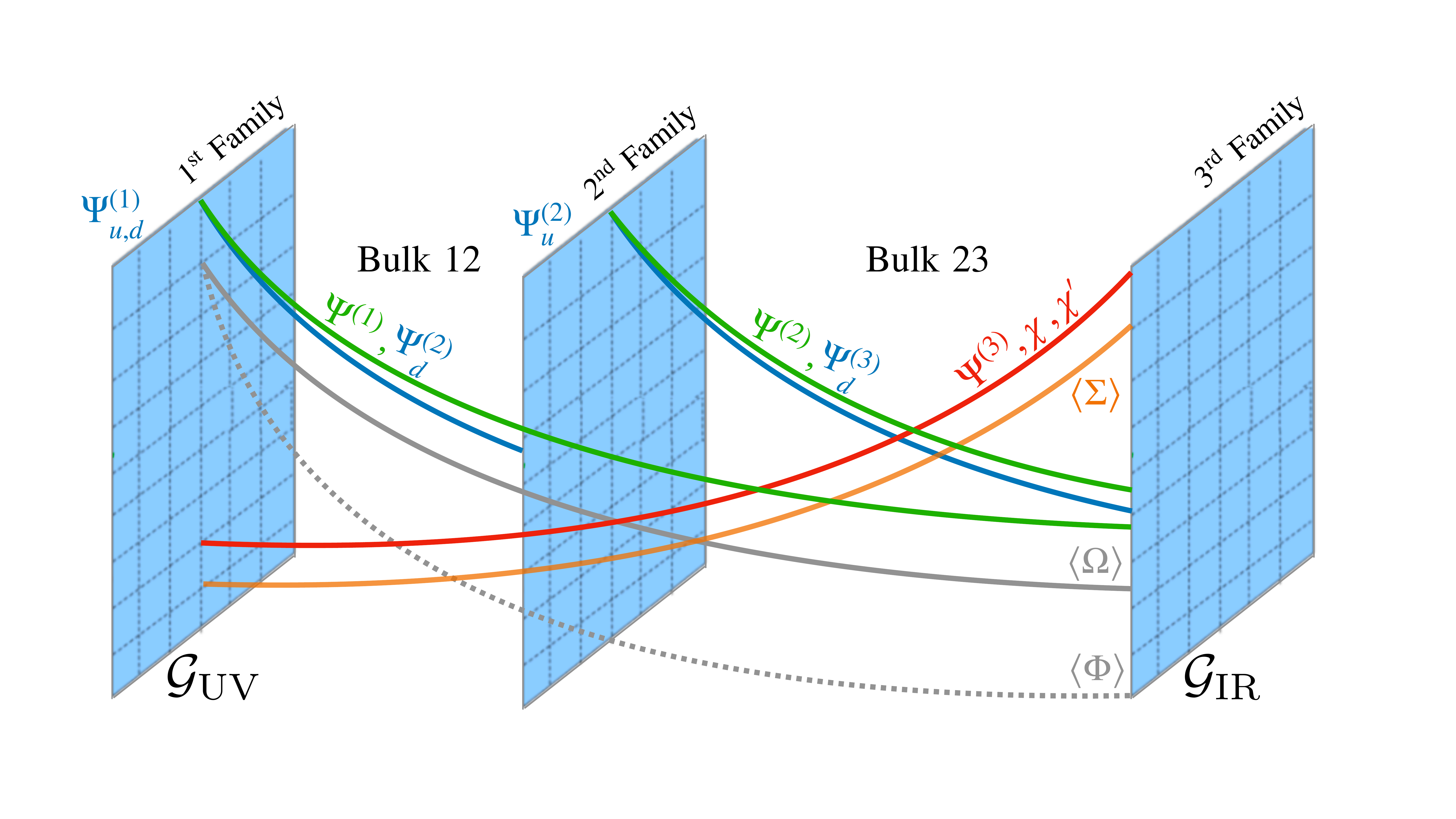}
\caption{Schematic representation of the 5D model. We include fermion zero-modes and scalar VEV profiles.}
\label{fig:3branes}
\end{figure}

We will now present a 5D model that combines all these elements commented above (for more technical details, see the original article\,\cite{Fuentes-Martin:2022xnb}): 
(i)~it stabilizes the Higgs scale à la RS,
(ii) below the KK scale, it reduces to a 4321 model that can explain the $B$-anomalies,
and (iii) it explains the flavor hierarchies from a multi-scale construction.
In Figure~\ref{fig:3branes} we illustrate the model.
The geometry thus consists on three branes (one for each family), with a compact extra dimension with the RS metric of Eq.~\eqref{RSgeometry}.\footnote{In general we can use different curvatures $k_{i}$ in the different bulks between the $i$-th brane and the $(i+1)$-th brane, with $k_i\leq k_{i+1}$ to avoid the appearance of ghosts. For simplicity here we will assume all curvatures are the same $k_i= k$.} To properly explain the flavor hierarchies, we need a distance between the first-family brane and the IR brane of $kL\sim 10$, and a first- to second-family brane distance of $k\ell \sim 4$.

On top of this, (iv)~the Higgs will be realized as a pseudo-Nambu-Goldstone boson (pNGB) of a global symmetry of the strongly coupled sector in the dual picture. In the 5D description, this corresponds to the gauge-Higgs unification, where the electroweak (EW) gauge group $SU(2)_L\times U(1)_Y$ is extended in the bulk to a larger group which is broken by boundary conditions on the branes. The fifth component of the gauge field associated to the broken generators both in the UV and IR can be associated to the Higgs.

\subsection{Gauge sector}

We will assume the following gauge groups on the 23 bulk and IR brane (see Figure~\ref{fig:3branes}),

\begin{align}\label{eq:GbulkIR}
\begin{aligned}
\mathcal{G}^{\rm 23}_{{\rm bulk}} &= SU(4)_h\times SU(3)_l\times U(1)_{B_l-L_l} \times  SO(5)\,, \\
\mathcal{G}_{\rm IR} &= SU(3)_c\times U(1)_{B-L}\times SO(4)\,.
\end{aligned}
\end{align}
The fifth component of the gauge fields associated to the broken generators in the IR brane behaves as Nambu-Goldstone bosons (NGBs) of a global symmetry $\mathcal{G}^{\rm 23}_{{\rm bulk}}$. The coset $\mathcal{G}^{\rm 23}_{{\rm bulk}}/\mathcal{G}_{\rm IR} $ contains 19 NGBs. Among them, 15 will be eaten by the $U_1$ leptoquark, the $Z^{\prime}$ and the color octet $G^{\prime}$ to get a mass $M_{15} \approx \Lambda_{\rm IR} \sqrt{2/(kL)}$. A good explanation of the $B$-anomalies is achieved taking $M_{15}\sim 3.5 - 4$\,TeV,\cite{Cornella:2021sby} so a good benchmark value is $\Lambda_{\rm IR}\sim 8\,$TeV. Such a value sets the first KK states, and first deviations of the 4321 model, for scales $\gtrsim 15\,$TeV. Deviations in EW precision data from these states is well below the current experimental limits.

The remaining 4 NGBs transform as a ${\bf 4}$ of $SO(4)=SU(2)_L\times SU(2)_R$ that we identify with the Higgs boson. This is similar to the minimal composite Higgs scenario based on the coset $SO(5)/SO(4)$.\cite{Agashe:2004rs} One advantage of realizing the Higgs as a pNGB is that we can compute its potential from other parameters of the model. It receives several contributions, at the tree and one-loop level from the matter content of the theory we discuss below. Assuming the value for the $SO(5)$ KK coupling $g_*=\sqrt{k}\,g_5\approx 2.5$, where $g_5$ is the 5D gauge coupling, all the contributions are calculated to be at the correct order. However, a fine-tuning at the per mille level is required to obtain a VEV for the Higgs at the EW scale.

There is some freedom on the gauge groups in the other branes and bulks, but we can choose
\begin{align}
\mathcal{G}^{\rm 2}_{{\rm brane}} &= \mathcal{G}^{\rm 23}_{{\rm bulk}}\,, \nonumber \\
\mathcal{G}^{\rm 12}_{{\rm bulk}} &= SU(4)_h\times SU(4)_l\times SO(5)\,, \nonumber \\
\mathcal{G}_{\rm UV} & = SU(4)_h\times SU(3)_l\times U(1)_{B_l-L_l}\times SU(2)_L\times U(1)_R\,.
\end{align}
We promote $SU(3)_l$ to $SU(4)_l$ in the 12 bulk so we unify quark and leptons of light families above the second-family brane scale. However, then the gauge group in the UV brane $\mathcal{G}_{\rm UV}$ needs to break $SU(4)_l$ to $SU(3)_l$ to explain the splitting of the first-family Yukawa couplings. Another possibility is to choose $\mathcal{G}^{\rm 12}_{{\rm bulk}} =\mathcal{G}^{\rm 2}_{{\rm brane}} = \mathcal{G}^{\rm 23}_{{\rm bulk}}$, and postpone the quark-lepton unification of light families to a more UV bulk we may include.

\subsection{Matter content}

The fermion and scalar fields in the model are given in Table~\ref{tab:content}, where we indicate their representation under the group $SU(4)_l\times SU(4)_h\times SO(5)$. The fields $\mathcal{S}_i$, $\Omega$ and $\Phi$ are necessary to implement an inverse seesaw mechanism to explain the neutrino masses and PMNS matrix.\cite{Fuentes-Martin:2020pww} The field $\Sigma$ is included to generate the light-family Yukawas as explained below. 
 The three scalar fields develop a VEV, and in particular, the VEV of $\Omega$ breaks the UV gauge symmetry $\mathcal{G}_{\rm UV}$ to the 4321 symmetry, $U(1)_{B_l-L_l}\times U(1)_R \to U(1)_X$.
The zero modes of the other fermion fields are exactly the 4321 fermion fields.  We have also included in Table~\ref{tab:content} the global groups $U(1)_{\Psi}$ and $U(1)_S$ we need to enforce, to forbid baryon and lepton number violation of higher dimensional operators and to realize the inverse seesaw mechanism, respectively. Alternatively, they may be gauge symmetries broken in a deeper Planck brane.

\begin{table}[t]
    \renewcommand{\arraystretch}{1.2}
    \centering
        \caption{Fermion and scalar content of the model. Here, $i=1,2,3$ and $j=1,2$. The upper block refers to fermion fields and the lower block to scalar fields.}
        \vspace{0.3cm}
    \begin{tabular}{|c|ccc||cc|}
    \hline
    Field & $SU(4)_h$ & $SU(4)_l$ & $SO(5)$ & $U(1)_\Psi$ & $U(1)_\mathcal{S}$ \\
    \hline
    \hline
    $\Psi^3,\Psi_d^3,\mathcal{X}^{({\prime})}$ & $\mathbf{4}$ & $\mathbf{1}$ & $\mathbf{4}$ & $1$ & $0$\\
    $\Psi^j,\Psi_{u,d}^j$        & $\mathbf{1}$ & $\mathbf{4}$ & $\mathbf{4}$ & $1$ & $0$\\
    $\mathcal{S}^i$         & $\mathbf{1}$ & $\mathbf{1}$ & $\mathbf{1}$ & $0$ & $1$\\
    \hline
    \hline
    $\Sigma$ & $\mathbf{1}$ & $\mathbf{1}$ & $\mathbf{5}$ & $0$ & $0$\\
    $\Omega$ & $\mathbf{1}$ & $\mathbf{4}$ & $\mathbf{4}$ & $1$ & $-1$\\ 
    $\Phi$   & $\mathbf{1}$ & $\mathbf{1}$ & $\mathbf{1}$ & $0$ & $2$\\ 
    \hline
    \end{tabular}
    \label{tab:content}
\end{table}

The boundary conditions in the UV and IR branes for the 4321 fermions are chosen to be $SU(4)_{l,h}$ symmetric, but $SO(5)$-breaking. They are
\begin{align}\label{eq:FermionBCs}
\Psi^3&=
\begin{bmatrix}
\psi^3\,(+,+)\\[2pt]
\psi_u^3\,(-,-)\\[2pt]
\tilde\psi_d^3\,(+,-)\\
\end{bmatrix}
\,,&
\Psi^3_d&=
\begin{bmatrix}
\tilde\psi^3\,(+,-)\\[2pt]
\tilde\psi_u^3\,(+,-)\\[2pt]
\psi_d^3\,(-,-)\\
\end{bmatrix}
\,,&
\mathcal{X}^{(\prime)}&=
\begin{bmatrix}
\chi^{(\prime)} (\pm,\pm)\\[2pt]
\chi^{(\prime)}_u\,(\mp,\pm)\\[2pt]
\chi^{(\prime)}_d\,(\mp,\pm)\\
\end{bmatrix}
\,,\nonumber\\
\Psi^j&=
\begin{bmatrix}
\psi^j\,(+,+)\\[2pt]
\tilde\psi_u^j\,(-,+)\\[2pt]
\tilde\psi_d^j\,(-,+)\\
\end{bmatrix}
\,,&
\Psi_u^j&=
\begin{bmatrix}
\tilde\psi^j\,(+,-)\\[2pt]
\psi_u^j\,(-,-)\\[2pt]
\hat\psi_d^j\,(+,-)\\
\end{bmatrix}
\,,&
\Psi_d^j&=
\begin{bmatrix}
\hat\psi^j\,(+,-)\\[2pt]
\hat\psi_u^j\,(+,-)\\[2pt]
\psi_d^j\,(-,-) \\
\end{bmatrix}\,,
\end{align}
where we are decomposing the ${\bf 4}$ of $SO(5)$ into the $SU(2)_L$ doublet, and the up-type and down-type components of the $SU(2)_R$ doublet. The sign $+$ ($-$) stands for a Dirichlet condition in the RH (LH) chirality of the fermion. Only the components with same boundary condition in the UV and IR have zero modes, with chirality LH for $(+,+)$ and RH for $(-,-)$.

\subsection{Yukawa hierarchies and U(2) breaking}

All the hierarchies can be explained choosing the appropriate 5D masses for the fermions. Different 5D masses will translate into different exponential profiles of the fermion zero modes. For concreteness, we can consider the following 5D masses $M_i=kc_i$:
\begin{align}\label{eq:cCoeff}
c_{\Psi^3}^{(1,2)}&\approx c_\mathcal{X}^{(1,2)} \approx c_{\Psi_d^3}^{(1)}\approx c_{\Psi^2}^{(1)}\approx c_{\Psi_u^2}^{(1)}\approx 0\,,~~~~~
c_{\mathcal{X}^\prime}^{(1,2)}\approx -1/2\,,\nonumber\\
c_{\Psi^1}^{(1,2)}&\approx c_{\Psi^2}^{(2)}\approx - c_{\Psi_d^3}^{(2)}\approx - c_{\Psi_d^2}^{(1)}\approx 1,~~~~~
c_{\Psi_{u,d}^1}^{(1,2)},~~c_{\Psi_{u,d}^2}^{(2)}\leq - 2\,,
\end{align}
where the superindex $(1,2)$ refers to the 12 or 23 bulk. The profiles generated are qualitatively depicted in Figure~\ref{fig:3branes}. This choice successfully reproduces all the hierarchies:

\begin{itemize}
\item  {\it Top Yukawa.} It is the only one directly generated in the bulk as a consequence that both $\psi^3$ ($SU(2)_L$ doublet) and $\psi^3_u$ (RH top) belong to the same multiplet $\Psi^3$. For $c_{\Psi^3}\approx 0$, $y_t \approx g_*/(2\sqrt{2})$, which is compatible with the prediction of $g_*$ from the Higgs potential.
\item {\it Down-type third-family Yukawas.}  The third family down-type Yukawa are generated through IR boundary masses we can write given the IR boundary conditions of the fields. The smallness of the bottom and tau Yukawa couplings versus the top one can be explained by localizing $\Psi_d^3$ (that contains $b_R$ and $\tau_R$) in the second brane ($c_{\Psi_d^3}^{(2)}\approx 1$). 
\item {\it Light-heavy Yukawa mixing.} The Yukawa couplings responsible of light-heavy mixing in the CKM matrix are also generated in the IR brane by boundary masses. They are suppressed by the profiles of $\Psi^j$, that contain the light-family LH doublets. By localizing the RH fields on the first- and second-family branes ($c_{\Psi_{u,d}^i}^{(2)}\leq - 2$), we suppress any breaking of the effective $U(2)$-symmetry at the TeV scale of the light RH sector in the light-heavy mixing.
\item {\it 4321 vector like fermions.} The IR masses also generate the mass of the vector like fermions and mass mixing with the SM fields, necessary to successfully explain the $B$-anomalies. Due to the profiles of the vector like fermions, $c_{\mathcal{X}^\prime}^{(1,2)}\approx -1/2$, and the SM fields, the mixing is maximal for the third family, and suppressed for the light families by the profile of the light-family LH fields $\Psi^j$. This is a similar suppression than the light-heavy Yukawa mixing.
\item {\it Light-family Yukawas.} Since RH fields of light families are very localized on the first- and second-family branes, light Yukawas can only be generated in those branes. In the Higgs-gauge unification scenario we are considering, we need to include the $\Sigma$ field, a ${\bf 5}$ of $SO(5)$ with a 5D mass close to the Breitenlohner-Freedman stability bound. It develops a VEV along its singlet component, propagating the breaking of $SO(5)$ into de bulk and the first- and second-family branes. The resulting Yukawa couplings are suppressed by the ratio between $\Lambda_{\rm IR}$ and the $i$-th brane scale, $\Lambda_{\rm IR}/\Lambda_{i}$,  for the light family $i$.
\end{itemize}

\section{Conclusions}

The anomalies observed in the $B$-meson decays hint New Physics at the TeV scale. A quark-lepton unification of the third family around the TeV scale, as in 4321 models, has a \-pheno\-me\-no\-logy consistent with the observed $B$-anomalies and the current experimental constraints. A~completion of 4321 models à la RS points towards a  multi-scale origin of the flavor hierarchies, where a $U(2)$ flavor protection at the TeV scale can be implemented.
In this talk we have presented a 5D model realizing this idea\,\cite{Fuentes-Martin:2022xnb} where, in addition, the Higgs emerges as a pNGB from the same dynamics that breaks the 4321 gauge symmetry to the SM group.

\section*{Acknowledgments}

JML would like to thank J. Fuentes-Martín, G. Isidori, B. A. Stefanek and N. Selimović for their valuable help in the preparation of this talk. This work is supported by the European Research Council (ERC) under the European Union’s Horizon 2020 research and innovation program under grant agreement 833280 (FLAY).

\end{document}